\documentclass[twocolumn]{jpsj2}
\usepackage{graphicx}

\def\runauthor{M.\ {\sc Eto}, T.\ {\sc Hayashi} and Y.\ {\sc Kurotani}}

\title{%
Spin Polarization at Semiconductor Point Contacts in \\
Absence of Magnetic Field
}

\author{%
Mikio {\sc Eto},\thanks{E-mail address: eto@rk.phys.keio.ac.jp}
Tetsuya {\sc Hayashi} and Yuji {\sc Kurotani}
}

\inst{%
Faculty of Science and Technology, Keio University, \\
3-14-1 Hiyoshi, Kohoku-ku, Yokohama 223-8522, Japan
}

\recdate{April 5, 2005}

\abst{%
Semiconductor point contacts can be a useful tool for producing
spin-polarized currents in the presence of spin-orbit (SO) interaction.
Neither magnetic fields nor magnetic materials are required.
By numerical studies, we show that
(i) the conductance is quantized in units of $2e^2/h$ unless the SO interaction
is too strong,
(ii) the current is spin-polarized in the transverse direction, and
(iii) a spin polarization of more than 50\% can be realized with experimentally
accessible values of the SO interaction strength. The spin-polarization
ratio is determined by the adiabaticity of the transition between subbands
of different spins during the transport through the point contacts.
}

\kword{%
point contact, Rashba, spin-orbit interaction, conductance quantization,
spin filter, Landau-Zener}

\begin{document}
\sloppy
\maketitle

Generating spin-polarized currents in semiconductors is an important issue for
the development of spin-based electronics, ``spintronics.''\cite{Zutic}
To manipulate electron spins, the Rashba spin-orbit (SO) interaction is useful
since its strength is locally controllable by applying an electric
field.\cite{Rashba,Nitta} Several
spin-filtering devices for producing the spin currents have been proposed utilizing
the SO interaction, {\it e.g}., three-terminal devices related to the spin
Hall effect,\cite{Kiselev,Pareek} a triple-barrier tunnel diode,\cite{Koga}
a one-dimensional system with a magnetic field,\cite{Streda}
and a three-terminal device for the Stern-Gerlach experiment using a nonuniform
SO interaction.\cite{Ohe}

In the present paper, we theoretically study the ballistic transport through a
semiconductor point contact (quantum wire with a narrow constriction)
in the presence of Rashba SO interaction. In its absence,
it is well known that the conductance is quantized in units of
$2e^2/h$ when the constriction changes gradually in
space.\cite{vanWees,Kawabata}
By numerical studies, we show that the conductance is quantized even with
the SO interaction and that the current is spin-polarized in the
transverse direction, in the absence of magnetic field. We
demonstrate that the polarization ratio can be more than 50\% in InGaAs
heterostructures. The spin current is obtained generally, $e.g$., in
GaAs heterostructures, if the condition later described by eq.\
(\ref{eq:condition}) is fulfilled
with not only Rashba but also Dresselhaus SO interaction.\cite{Dresselhaus}
As spin filters, two-terminal devices with point contacts are easy
to fabricate on semiconductors, compared with other devices that have been
proposed.\cite{Kiselev,Pareek,Koga,Streda,Ohe}

We consider a two-dimensional electron gas confined in the $z$ direction.
The electric field in the $z$ direction results in the Rashba SO interaction,
\begin{equation}
H_{\rm RSO}=\frac{\alpha}{\hbar}(p_y\sigma_x-p_x\sigma_y),
\end{equation}
where $\sigma_x$ and $\sigma_y$ are Pauli matrices.
(The Dresselhaus SO interaction is discussed later.)
We use a dimensionless parameter, $k_{\alpha}/k_{\rm F}$, where
\begin{equation}
k_{\alpha}=m \alpha/\hbar^2
\end{equation}
with $m$ being the effective mass and $k_{\rm F}$ is the Fermi
wavenumber.\cite{com1} In InGaAs heterostructures,
$\alpha=(3 \sim 4) \times 10^{-11}$ eVm and
$\Delta_{\rm R} \equiv 2k_{\rm F}\alpha=15 \sim 20$ meV
[$k_{\alpha}/k_{\rm F}=\Delta_{\rm R}/(4E_{\rm F}) \approx
0.1$].\cite{Grundler,Yamada}
Electron-electron interaction and impurity scattering are neglected.
Electrons propagate in a quantum wire along the $x$ direction, with
width $W_0$ in the $y$ direction. A hard-wall confinement potential $U(y)$
is assumed, $U(y)=0$ for $-W_0/2<y<W_0/2$ and $\infty$ otherwise.
For a narrow constriction around $x=y=0$, we consider an extra potential
at $-L_1<x<L_2$, which is analogous to that adopted in ref.\ 15:
%\cite{Ando}:
%\begin{eqnarray}
%V(x,y) & = & \frac{V_0}{2} \left( 1+\cos\frac{\pi x}{L_x} \right)
%\nonumber \\
%& & +E_{\rm F} \sum_{\pm} \left[ \frac{y-y_{\pm}(x)}{\Delta} \right]^2
%\theta(\pm[y-y_{\pm}(x)]),
%\label{eq:model}
%\end{eqnarray}
\[ %\begin{eqnarray}
V(x,y) = \frac{V_0}{2} \left( 1+\cos\frac{\pi x}{L_x} \right)
%\nonumber \\
\]
\begin{equation}
 +E_{\rm F} \sum_{\pm} \left[ \frac{y-y_{\pm}(x)}{\Delta} \right]^2
\theta(\pm[y-y_{\pm}(x)]),
\label{eq:model}
\end{equation}
with
\begin{equation}
y_{\pm}(x)=\pm\frac{W_0}{4} \left( 1-\cos\frac{\pi x}{L_x} \right),
\end{equation}
where $L_x=L_1$ ($-L_1<x<0$) and $L_x=L_2$ ($0<x<L_2$).
$\theta(t)$ is a step function
[$\theta(t)=1$ for $t>0$, $0$ for $t<0$].
For fixed $x$, $V(x,y)$ is flat at $y_-(x)<y<y_+(x)$ and grows with
increasing $|y-y_{\pm}(x)|$ at $y>y_+(x)$ or $y<y_-(x)$. On a line of $y=0$,
the potential height is given by $(V_0/2) [1+\cos (\pi x/L_x)]$, being
maximal at $x=0$. In the present paper,
we fix $W_0=4\lambda_{\rm F}$ and $\Delta=\lambda_{\rm F}$, where
$\lambda_{\rm F}$ is the Fermi wavelength ($\lambda_{\rm F}=2\pi/k_{\rm F}$).

Numerical calculations are performed using a tight-binding model on a square
lattice ($-L_1<x<L_2$, $-W_0/2<y<W_0/2$), following refs.\
%\cite{Ando,Tamura}
15 and 16.\cite{com0}
The transmission coefficients are evaluated for incident electrons
from the left side of the constriction ($x<-L_1$) to the right ($L_2<x$),
using the Green function's recursion method.\cite{Ando}
They yield the conductance $G$ through the Landauer formula.

%------------------------ Fig.1------------------------
\begin{figure}[hbt]
\begin{center}
\includegraphics[width=7cm]{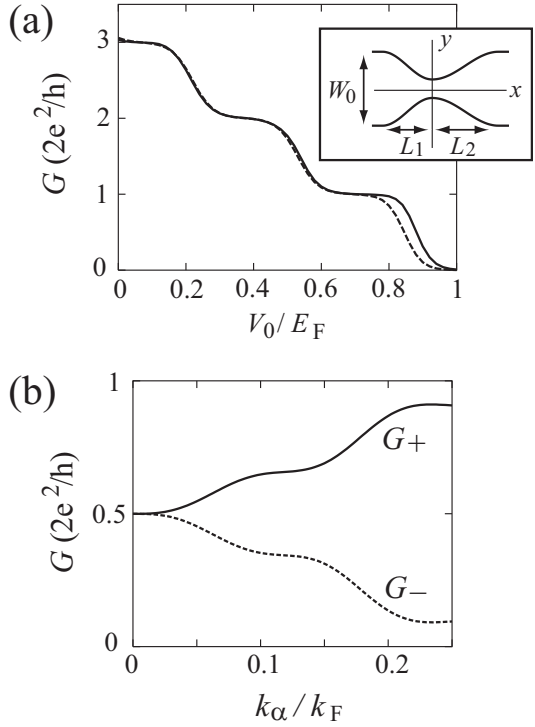}
\end{center}
\caption{Numerical results of the conductance $G$ through a point contact
with $L_1=L_2=4 \lambda_{\rm F}$.
(a) $G$ as a function of $V_0$, potential height at $x=y=0$.
The strength of the spin-orbit (SO) interaction is
$k_{\alpha}/k_{\rm F}=0.25$ (solid line). The broken line indicates $G$ in
the absence of SO interaction. Inset: Schematic drawing of our model
($W_0=4\lambda_{\rm F}$).
(b) The conductance $G_{\pm}$ for electrons with $S_y=\pm 1/2$ in the output
current, as a function of $k_{\alpha}/k_{\rm F}$.
$V_0=0.7E_{\rm F}$ where $G$ is at the first plateau ($G=G_++G_- = 2e^2/h$).}
\end{figure}

Figure 1 shows the calculated results when $L_1=L_2=4\lambda_{\rm F}$.
In Fig.\ 1(a), we plot $G$ as a function of $V_0$, potential height
at $x=y=0$. The conductance quantization is clearly
observed when $k_{\alpha}/k_{\rm F}=0.25$ (solid line)
as well as in the absence of SO interaction (broken line).
The conductance quantization is broken for
$k_{\alpha}/k_{\rm F} > 0.5$ (not shown here),
which is addressed later. We divide the output current
into two components, one carried by spin-up electrons in the $y$
direction ($S_y=1/2$) and the other by spin-down electrons ($S_y=-1/2$).
Figure 1(b) presents each conductance, $G_{\pm}$, as a function of 
$k_{\alpha}/k_{\rm F}$. $V_0=0.7E_{\rm F}$
where the total conductance is at the first plateau ($G=G_++G_-=2e^2/h$).
The spin polarization in the $y$ direction, $(G_+-G_-)/(G_++G_-)$,
increases with an increase in $k_{\alpha}/k_{\rm F}$.
Note that incident electrons are unpolarized in eight conduction
channels per spin direction.
These results indicate that (i) the point contact works
as a spin filter, (ii) the polarization ratio is about 30\% with experimental
values of $k_{\alpha}/k_{\rm F} \approx 0.1$ in this case, and
(iii) the conductance is still quantized.

These calculated results can be understood as follows.
We divide the Hamiltonian into two parts: $H=H_0+H'$,
\begin{eqnarray}
H_0 & = & \frac{1}{2m}(p_x^2+p_y^2)-\frac{\alpha}{\hbar} p_x \sigma_y
+V(x,y)+U(y), \\
H' & = & \frac{\alpha}{\hbar} p_y \sigma_x.
\label{eq:Hprime}
\end{eqnarray}
We treat $H'$ as a perturbation. When $V(x,y)$ is independent of $x$,
the eigenstates and eigenvalues of $H_0$ are given by
\begin{eqnarray}
\psi_{n,k,\pm} & = & e^{ikx}\varphi_n(y)\chi_{\pm}, \\
E_{n,\pm}(k) & = & \frac{\hbar^2}{2m}(k \mp k_{\alpha})^2-
\frac{\hbar^2}{2m}k_{\alpha}^2+\varepsilon_n,
\end{eqnarray}
respectively, where $\varphi_n(y)$ are eigenstates of $p_y^2/(2m)+V(y)+U(y)$
with eigenvalues $\varepsilon_n$ ($n=1,2,3,\cdots$).
$\chi_{\pm}$ are eigenstates of $\sigma_y$, representing spin-up or
-down states.
The dispersion relations of $E_{n,\pm}(k)$ (subbands) are schematically
shown in Figs.\ 2(a) and 2(b). The Fermi level is denoted by horizontal
line $A$ in the quantum wire outside of the constriction,
where $V(x,y)=0$.
(The number of channels is three per spin direction in the figures, which
is smaller than that in our numerical studies.)
There are two situations regarding the crossings between
$E_{n,-}(k)$ and $E_{n',+}(k)$ ($n<n'$). When
$k_{\alpha}/k_{\rm F}<(3/16)(\lambda_{\rm F}/W_0)^2$,
all the crossings take place above $E_{\rm F}$ [Fig.\ 2(a)]. When
\begin{equation}
k_{\alpha}/k_{\rm F}>(3/16)(\lambda_{\rm F}/W_0)^2,
\label{eq:condition}
\end{equation}
some of them appear below $E_{\rm F}$ [Fig.\ 2(b)].\cite{com2}

%------------------------ Fig.2------------------------
\begin{figure}[t]
\begin{center}
\includegraphics[width=8cm]{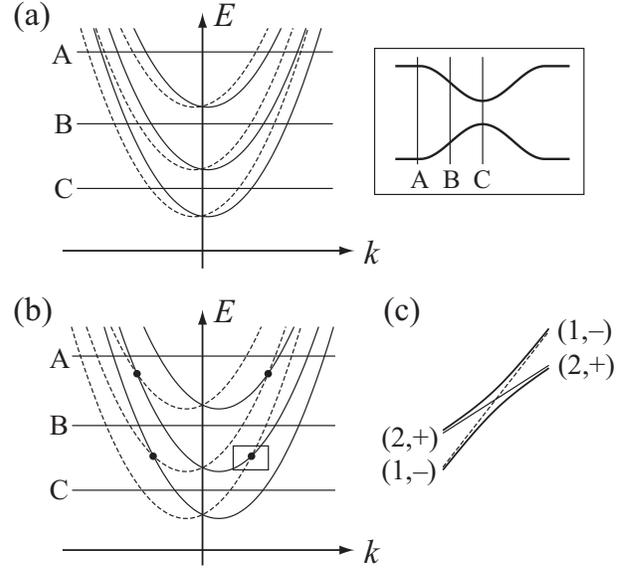}
\end{center}
\caption{Schematic drawings of subbands, $E_{n,+}(k)$ (solid lines)
and $E_{n,-}(k)$ (broken lines) with $n=1,2,3$, in a quantum wire with
a component of SO interaction, $-(\alpha/\hbar) p_x \sigma_y$.
All the crossings between $E_{n,-}(k)$ and
$E_{n',+}(k)$ ($n<n'$) take place above the Fermi energy $E_{\rm F}$ in
case (a), whereas some of them appear below $E_{\rm F}$ in case (b).
Horizontal lines indicate $E_{\rm F}$ (relative to the subbands)
corresponding to three positions of electrons in the point contact,
shown in the inset in (a).
(c) A vicinity of the crossing between $E_{1,-}(k)$ and $E_{2,+}(k)$, which
is surrounded by a square in (b). The subbands are mixed with each other by
the other component of SO interaction, $(\alpha/\hbar) p_y \sigma_x$.}
\end{figure}

Now we consider the transport of electrons through a constriction
when the conductance shows the first plateau.
Except in the vicinities of the above-mentioned crossings,
we assume an adiabatic transport in which the wavefunction $\psi_{n,k,\pm}$
changes gradually remaining the quantum numbers of transverse motion $n$ and
of spin $\pm$.\cite{vanWees,Kawabata} The wavenumber $k$ changes with $x$,
which is determined by an intersection between the subband and $E_{\rm F}$;
positive $k$'s for incident electrons.
Let us begin with the case of Fig.\ 2(a).
As electrons propagate from the wire to a narrow
constriction ($x<0$), the subbands shift upwards. Alternatively, we move
$E_{\rm F}$ downwards in Fig.\ 2(a). Similarly, at $x>0$, we move
$E_{\rm F}$ upwards. [Separations between the subbands increase (decrease)
with $x$ at $x<0$ ($x>0$), which is not shown in the figure.]
Before the injection into the constriction (position A), there are six
conduction modes, $(1,\pm)$, $(2,\pm)$ and $(3,\pm)$. At position B,
modes $(1,\pm)$ and $(2,\pm)$ are conducting, whereas modes $(3,\pm)$ have
been completely reflected. At the narrowest region (position C),
only modes $(1,\pm)$ exist. At $x>0$, the modes propagate with the
transmission probability of unity, which results in the conductance
quantization, $G=2e^2/h$.
The small perturbation of $H'$ does not play an important role.
No spin polarization is observed in this case.

In the case of Fig.\ 2(b), the situation is different.
Electrons pass by the crossings twice, once at $x<0$ and
once at $x>0$. Let us look at the crossing between modes $(1,-)$ and $(2,+)$.
These subbands are mixed by the perturbation $H'$, as shown in
Fig.\ 2(c). Hence these modes change to each other with a transition
probability $P$ when electrons pass through the crossing. Around the first
pass ($x<0$), both modes are occupied by electrons
just before the modes cross.
Then spin-up electrons are flipped to spin-down with probability $P$,
while spin-down electrons are flipped to spin-up with the same probability.
Accordingly, no spin polarization takes place.
Around the second pass ($x>0$), on the other hand, mode $(2,+)$ is empty
while mode $(1,-)$ is full of electrons just before the modes cross.
Then spin-down electrons in the latter mode are
spin-flipped to spin-up in the former mode with probability $P$.
The spin-up electrons in mode $(1,+)$ are transmitted through the constriction
without passing by any mode crossing. Consequently we obtain the
spin-polarization ratio of $[(1+P)-(1-P)]/2=P$.

%------------------------ Fig.3------------------------
\begin{figure}[t]
\begin{center}
\includegraphics[width=6cm]{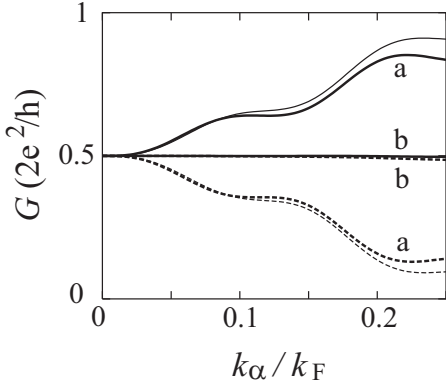}
\end{center}
\caption{Conductance $G_{\pm}$ for electrons with $S_y=\pm 1/2$ in
the output current, as a function of $k_{\alpha}/k_{\rm F}$ (solid lines
for $G_+$ and broken lines for $G_-$).
$V_0=0.7E_{\rm F}$ in eq.\ (\ref{eq:model}) where $G$ is at the first
plateau ($G=G_++G_- = 2e^2/h$). $L_1=L_2=4 \lambda_{\rm F}$.
A term of $(\alpha/\hbar) p_y \sigma_x$ in $H_{\rm RSO}$
is taken into account (a) only at $x>0$ or
(b) only at $x<0$, whereas the other term, $-(\alpha/\hbar) p_x \sigma_y$,
is considered in the whole region.
The thin lines indicate $G_{\pm}$ when both the terms are present
in the whole region.}
\end{figure}

The transition probability $P$ is evaluated using the Landau-Zener
theory:\cite{Landau,Zener}
\begin{equation}
P=1-\exp(-2\pi\lambda),
\label{eq:LZ}
\end{equation}
where $\lambda=J^2/(\hbar |v|)$ represents the degree of adiabaticity.
$J=|\langle 2,+| H' | 1,- \rangle |$ and $v=\frac{\partial}{\partial t}
[E(2,+)-E(1,-)]$, the velocity of the change of level spacing. $P=1$ for
$\lambda=\infty$ in the adiabatic limit, whereas $P=0$ for $\lambda=0$
in the sudden-change limit. If $V(x,y)$ is a hard-wall potential with
width $W(x)$ in the $y$ direction, instead of by
eq.\ (\ref{eq:model}), $\lambda$ is estimated as
\begin{equation}
\lambda=k_{\alpha} \cdot \left| \frac{1}{W} \frac{dW}{dx} \right|^{-1},
\label{eq:lambda}
\end{equation}
apart from a numerical factor. $W$ and its derivative should be evaluated
at $x$ where electrons pass by the crossing.
In more gradual point contacts (smaller $|\frac{1}{W} \frac{dW}{dx}|$),
electrons pass through the crossing more adiabatically (larger $\lambda$).
Then the larger spin-flip probability $P$, and thus the larger
spin-polarization ratio, is expected.

In spite of the spin polarization, the total conductance is not affected
by the mode crossings, unless the SO interaction is too strong.
In Fig.\ 2(b), both the derivatives of $E_{1,-}(k)$ and $E_{2,+}(k)$
are positive at their crossing. Since the group velocities have the same
sign, the transition from one mode to the other is not accompanied by
a reflection (forward scattering).
If the SO interaction were so strong that the crossing occurred at
$k<k_{\alpha}$, backward scattering could take place,
which would destroy the conductance quantization.
The condition that the backward scattering does not take place is
given by $2\hbar^2 k_{\alpha}^2/m < \varepsilon_2-\varepsilon_1$, or
\begin{equation}
k_{\alpha}/k_{\rm F} < (\sqrt{3}/4)(\lambda_{\rm F}/W)
\label{eq:conditionb}
\end{equation}
for the hard-wall confinement of width $W(x)$. In eq.\
(\ref{eq:conditionb}), $W$ is the width of confinement where modes
$(1,-)$ and $(2,+)$ cross ($\lambda_{\rm F} < W < W_0$). [In our
numerical studies, this condition seems to be satisfied with
$k_{\alpha}/k_{\rm F}<0.25$. We could still observe a spin-polarized current
when eq.\ (\ref{eq:conditionb}) does not hold, as discussed later.]

%------------------------ Fig.4------------------------
\begin{figure}[t]
\begin{center}
\includegraphics[width=6.5cm]{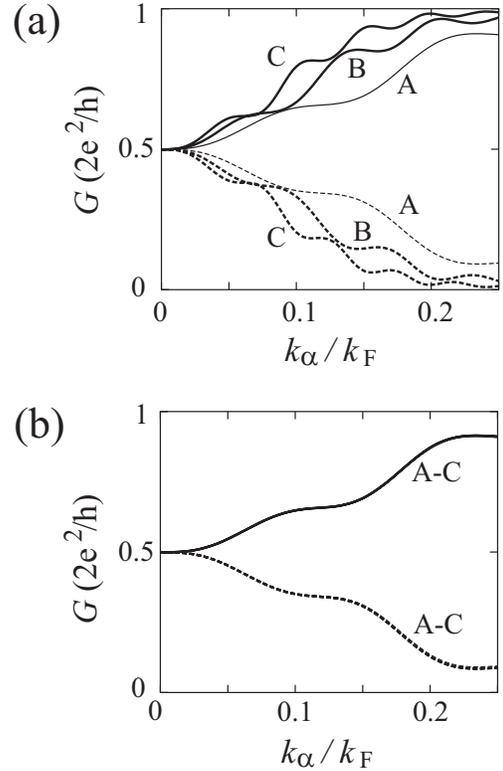}
\end{center}
\caption{Conductance $G_{\pm}$ for electrons with $S_y=\pm 1/2$ in
the output current, as a function of $k_{\alpha}/k_{\rm F}$ (solid lines
for $G_+$ and broken lines for $G_-$).
$V_0=0.7E_{\rm F}$ in eq.\ (\ref{eq:model}) where $G$ is at the first
plateau ($G=G_++G_- = 2e^2/h$).
(a) $L_1$ is fixed at $4 \lambda_{\rm F}$, whereas $L_2$ is changed
as (A) $4\lambda_{\rm F}$, (B) $8\lambda_{\rm F}$ and
(C) $12\lambda_{\rm F}$.
(b) $L_2$ is fixed at $4 \lambda_{\rm F}$, whereas $L_1$ is changed
as (A) $4\lambda_{\rm F}$, (B) $8\lambda_{\rm F}$ and
(C) $12\lambda_{\rm F}$.
}
\end{figure}

We have so far presented a simple theory to explain the numerical results
in Fig.\ 1, on the basis of the adiabatic approximation for the transport
through the point contact and perturbative treatment of $H'$ in eq.\
(\ref{eq:Hprime}). To verify this theory, we further perform numerical studies
and present the results in Figs.\ 3 and 4.

In our theory, the spin-flip by $H'$ is important for the spin polarization
during the transport from a narrow region to a wide region ($x>0$).
To confirm this idea, we examine two situations in Fig.\ 3:
(a) $H'$ is present only at $x>0$ or (b) only at $x<0$.
Indeed we observe the spin polarization in situation (a), whereas the
polarization is not visible in situation (b).

The adiabaticity of the spin-flip transition can be controlled by changing
the shape of point contacts, {\it e.g}., $(L_1,L_2)$ in our model.
With increasing $L_2$,
$|\frac{1}{W} \frac{dW}{dx}|$ decreases in eq.\ (\ref{eq:lambda}) and
hence $P$ in eq.\ (\ref{eq:LZ}) increases. As a result, a larger
spin-polarization is expected.
Figure 4(a) shows the numerical results with
(A) $(L_1,L_2)=(4\lambda_{\rm F},4\lambda_{\rm F})$,
(B) $(4\lambda_{\rm F},8\lambda_{\rm F})$, and
(C) $(4\lambda_{\rm F},12\lambda_{\rm F})$. The polarization ratio
increases with an increase in $L_2$, in accordance with our theory.
In case (C), we observe a
polarization of 60\% with experimental values of $\alpha$ in
InGaAs heterostructures.
In Fig.\ 4(b), we change $L_1$ with a fixed $L_2$.
The spin polarization is not influenced by $L_1$, which is consistent with
the previous discussion.

In conclusion, we have examined the ballistic transport through semiconductor
point contacts in the presence of Rashba SO interaction.
We have observed a spin-polarized current although the conductance is
still quantized. The spin-polarization ratio is
determined by the adiabaticity of the transition between subbands of
different spins, which is characterized by eq.\ (\ref{eq:lambda}), during
the transport from a narrow region to a wide region.

We have examined a quantum wire of width $W_0=4\lambda_{\rm F}$ with
a narrow constriction and
demonstrated that the polarization ratio can be 60\% in InGaAs
heterostructures. Generally, a condition to generate the spin
current is given by eq.\ (\ref{eq:condition}).
In GaAs heterostructures, $W_0>6\lambda_{\rm F}$ is required with
$\alpha=0.05 \times 10^{-11}$ eVm
($k_{\alpha}/k_{\rm F} \sim 0.005$).\cite{GaAs}
Our mechanism also works with Dresselhaus SO interaction
\begin{equation}
H_{\rm DSO}=\frac{\alpha'}{\hbar}(-p_x\sigma_x+p_y\sigma_y).
\end{equation}
Then the output current is spin-polarized in the $x$ direction.
If $H_{\rm RSO}$ and $H_{\rm DSO}$ coexist,\cite{GaAs} the eigenstates of
$\alpha \sigma_y + \alpha' \sigma_x$ determine the direction of
spin polarization.

We make a few remarks. (i) The spin filtering effect is expected at the
higher plateaus of conductance as well as at the first plateau.
At the second plateau, for example, the crossings between modes $(1,-)$ and
$(n,+)$ and those between $(2,-)$ and $(n,+)$ ($n>2$) work for the spin
polarization if they appear below $E_{\rm F}$. Note that the crossing
between modes $(1,-)$ and $(2,+)$ does not work since both modes are
occupied just before the modes cross even at the second time.
(ii) A steplike structure of the spin polarization is seen as a function
of $\alpha$ in Figs.\ 1(b) and 4. This reflects the number of
crossings between $(1,-)$ and $(n,+)$ with $n>1$ below $E_{\rm F}$.
With an increase in $\alpha$, more crossings appear below $E_{\rm F}$,
which increases the spin-flip probability.
(iii) In our model, the conductance seems to be influenced by the
backward scattering for $k_{\alpha}/k_{\rm F}>0.5$ [eq.\ (\ref{eq:conditionb})
does not hold]. Around $k_{\alpha}/k_{\rm F}=0.7$, we observe the
conductance plateaus with a small fluctuation, which might be due to
some resonant states around the point contact with backward scattering.
A spin current is still obtained with a polarization ratio of $\sim$90\%.
When the SO interaction is increased further, the plateaus begin to break.
The investigation of this regime in more detail is beyond the scope of
this paper.

Finally, we discuss the observation of the spin-polarized current produced
by our mechanism. The spin polarization has to be directly measured since the
conductance is not influenced. Possible experiments are
an injection of the spin current into dilute magnetic semiconductors,
an injection into a spin detector,\cite{Folk,Elzerman} or
an optical measurement.\cite{Kato}
An indirect measurement may be available owing to the fact that mode $(1,-)$
is converted to upper modes, $(n,+)$ with $n>1$, in the spin-flip processes.
If the obtained current is injected into another point contact in the
absence of SO interaction, the height of the first plateau is suppressed
to $\sim (2-P) e^2/h$ when the polarization ratio is $P$.

%----------------------------
The authors gratefully acknowledge discussions with D.\ Matsubayashi,
H.\ Yokouchi, J.\ Yamauchi, Y.\ Tokura and G.\ E.\ W.\ Bauer.
This work was partially supported by a Grant-in-Aid for
Scientific Research in Priority Areas ``Semiconductor Nanospintronics''
(No.\ 14076216) of the Ministry of Education, Culture, Sports, Science
and Technology, Japan.


\begin{thebibliography}{99}
\bibitem{com1}
$k_{\rm F}$ ($E_{\rm F}$) is the Fermi wavenumber (energy) in the
absence of SO interaction; $E_{\rm F}=\hbar^2k_{\rm F}^2/(2m)$.
In its presence, we shift the Fermi energy by
$-\hbar^2k_{\alpha}^2/(2m)$ to keep the electron density constant.
\bibitem{com0}
The lattice constant $a$ is chosen as $\lambda_{\rm F}/a=8$ when
$k_{\alpha}/k_{\rm F}<0.25$. A smaller value is required for $a$
when $k_{\alpha}/k_{\rm F}>0.25$.
\bibitem{com2}
Equation (\ref{eq:condition}) is an approximate expression that is 
valid for $k_{\rm F}^2 \gg (2\pi/W_0)^2$. It is exactly given by
$k_{\alpha}
[\sqrt{k_{\rm F}^2-(\pi/W_0)^2}+\sqrt{k_{\rm F}^2-(2\pi/W_0)^2}]/2
>(3/4)(\pi/W_0)^2$.
\end{thebibliography}
\end{document}